\documentclass{article}
\usepackage{PRIMEarxiv}
\usepackage[T1]{fontenc}    
\usepackage{hyperref}       
\usepackage{url}            
\usepackage{booktabs}       
\usepackage{amsfonts}       
\usepackage{nicefrac}       
\usepackage{microtype}      
 \usepackage{graphicx} 
 \graphicspath{{images/}} 
\usepackage{lipsum}
\usepackage{amsmath}
\usepackage{listings}
\usepackage{xltabular, booktabs}
\usepackage{xcolor}
\lstdefinelanguage{Solidity}{
    keywords={function, public, view, returns, require, emit, uint256, bool, address, mapping},
    keywordstyle=\color{blue}\bfseries,
    ndkeywords={block, timestamp, msg, sender},
    ndkeywordstyle=\color{purple}\bfseries,
    identifierstyle=\color{black},
    sensitive=false,
    comment=[l]{//},
    morecomment=[s]{/*}{*/},
    commentstyle=\color{green}\ttfamily,
    stringstyle=\color{red}\ttfamily,
    morestring=[b]',
    morestring=[b]"
}
\definecolor{keywordblue}{RGB}{0,0,255}
\definecolor{commentgreen}{RGB}{0,128,0}
\definecolor{stringred}{RGB}{163,21,21}
\definecolor{background}{RGB}{245,245,245}

\lstdefinelanguage{JavaScript}{
    keywords={let, await, async, for, while, if, try, catch, throw, else, console, function, return},
    keywordstyle=\color{keywordblue}\bfseries,
    ndkeywords={deployed, getAccounts, isSuccess, mineBlock, getMinerBlockNumbers, getTotalBlockCount, toString},
    ndkeywordstyle=\color{keywordblue}\bfseries,
    commentstyle=\color{commentgreen}\ttfamily,
    stringstyle=\color{stringred}\ttfamily,
    numbers=left,
    numberstyle=\tiny\color{gray},
    frame=single,
    breaklines=true,
    showstringspaces=false
}

\usepackage{booktabs}
\usepackage{enumitem} 
\usepackage{lastpage}   
\usepackage{fancyhdr}       
\usepackage{graphicx}       
\graphicspath{{media/}}     
\usepackage{booktabs} 
\usepackage{longtable}
\usepackage{indentfirst} 
\usepackage{indentfirst} 
\usepackage{tabularx} 
\graphicspath{{images/}}
\usepackage{titlesec}
\titleformat{\subsubsection}
  {\normalfont\bfseries\itshape} 
  {\thesubsubsection}            
  {1em}                          
  {}                             

\titlespacing{\subsubsection}
  {0pt}     
  {1ex}     
  {0.5ex}   
\usepackage{float}
\usepackage{listings}
\usepackage{xcolor}
\fancypagestyle{plain}{%
  \fancyhf{\headrule} 
\fancyfoot[LO]{Xiaoyan Zhang et.al.: \itshape Preprint Submitted to Arxiv.\\ \textsuperscript{*}These authors contributed equally to this work. }
\fancyfoot[Ro]{Page \thepage\ of \pageref{LastPage}} 
}

\title{Risk Assessment and Security Analysis of Large Language Models}

\author{
  Xiaoyan Zhang\textsuperscript{*}\\
  School of Cyberspace Security\\
  Hainan University \\
  \texttt{xia0yanzhang@outlook.com} \\
   \And
     Dongyang Lyu\textsuperscript{*} \\
  School of Cyberspace Security\\
  Hainan University \\
  \texttt{dongyanglyu@hainanu.edu.cn} \\
  \And
  Xiaoqi Li \\
 School of Cyberspace Security \\
  Hainan University \\
  \texttt{csxqli@ieee.org} \\
}

\begin{document}
\maketitle 
\thispagestyle{plain}
\begin{abstract}
As large language models (LLMs) expose systemic security challenges in high risk applications, including privacy leaks, bias amplification, and malicious abuse, there is an urgent need for a dynamic risk assessment and collaborative defence framework that covers their entire life cycle. This paper focuses on the security problems of large language models (LLMs) in critical application scenarios, such as the possibility of disclosure of user data, the deliberate input of harmful instructions, or the models bias. To solve these problems, we describe the design of a system for dynamic risk assessment and a hierarchical defence system that allows different levels of protection to cooperate. This paper presents a risk assessment system capable of evaluating both static and dynamic indicators simultaneously. It uses entropy weighting to calculate essential data, such as the frequency of sensitive words, whether the API call is typical, the realtime risk entropy value is significant, and the degree of context deviation. The experimental results show that the system is capable of identifying concealed attacks, such as role escape, and can perform rapid risk evaluation. The paper uses a hybrid model called BERT-CRF (Bidirectional Encoder Representation from Transformers) at the input layer to identify and filter malicious commands. The model layer uses dynamic adversarial training and differential privacy noise injection technology together. The output layer also has a neural watermarking system that can track the source of the content. In practice, the quality of this method, especially important in terms of customer service in the financial industry. We find that it is much faster to handle new attacks, about three times faster than the previous solution. Most importantly, although the speed is faster, the quality of the text generated is still as good as before and unaffected. We experimented with NVIDIA A100 GPU clusters and real industry datasets, and it turned out that the framework worked. The analysis found that while the defence overhead of hyperscale models grew more slowly and not as fast as linear, they could do better in defending against multimodal attacks. The approach proposed in this paper provides a viable technical solution to LLM security governance, which can be used in multimodal learning and colearning in these other application scenarios.
\end{abstract}


\section{INTRODUCTION}
Large Language Models (LLMs) based on the Transformer architecture achieve capability evolution through multi stage training strategies. Initially, they are parameterised using massive datasets such as OpenWebText and Common Crawl. Subsequently, by continuously expanding parameter scales as seen in models like the GPT series, PaLM, and LLaMA, they progressively expand application boundaries, achieving a paradigm shift from basic text generation to multimodal reasoning. Research indicates that these models have demonstrated practical value in vertical fields such as medical diagnosis assistance and financial risk prediction. However, the risks exposed in their actual application also require attention. Specifically, the exponential growth of model parameters to the trillionlevel scale and the widespread deployment of open interfaces collectively present three security challenges: the risk of training data privacy leaks, the amplification of implicit biases in model decision making, and the possibility of maliciously abused generated content. For example, in adversarial tests on GPT-4, the success rate of jailbreak attack samples in laboratory environments exceeded one quarter\cite{vaswani2017attention}. Empirical research on the CivilComments dataset indicates that user comments involving minority groups exhibit systematic emotional bias, with the probability of negative emotion labelling being 15\% to 20\% higher than the baseline value. The iterative upgrading of member inference attacks and training data extraction attacks indicates that security threats to large language models have expanded from traditional user data leaks to the development of automated attack tools in the technical black market. An urgent need is to establish a dynamic risk assessment mechanism with multi layered defensive features such as privacy protection, bias reduction, and abuse prevention\cite{brown2020language}. This is an intrinsic requirement of AI technology evolution and a key pathway to addressing the ethical challenges of deploying intelligent systems in society.

This study develops a dynamic risk perception | layered defence response collaborative mechanism, which integrates static and dynamic indicators to establish a multi-dimensional risk assessment system \cite{yao2024survey}\cite{shi2024challenges}. Regarding indicator weight allocation, the entropy weight method is used for objective weighting, providing a technical contrast to traditional static assessment methods\cite{cui2024risk}. Based on this framework, the Prometheus real time monitoring API is integrated to build a risk dynamic quantification system, and the NSFOCUS Risk Matrix v1 is used to develop an adaptive threshold partitioning algorithm. Through dynamic mapping of risk levels (T1 to T4), this algorithm can effectively detect new attack patterns such as Jailbreaking Attacks and Role Play Escape, with detection latency compressed to within 50ms \cite{derczynski2024garak}.

The existing experimental verification system has limitations in terms of dynamic threat coverage \cite{ganguli2022red}\cite{derczynski2024garak}. Based on this, this study designed and implemented a multi source hybrid verification scheme that integrates three data sources: public datasets, GPT-4 synthetic data, and industry specific anonymised data. This scheme establishes an experimental environment with dynamic evolutionary characteristics. Specifically, the experimental scenarios cover complex attack patterns such as zero sample jailbreak attacks and cross language prompt injection attacks. The TextAttack tool was used to generate attack samples, with an initial attack success rate consistently maintaining the 25\% baseline \cite{ganguli2022red}. This setting effectively evaluates the adaptability of layered defence mechanisms against real time evolving adversarial attacks.
The experimental data indicate that the defence mechanisms perform well across multiple dimensions. In terms of security protection, the success rate of malicious code execution decreased from the baseline value of 23\% to 1\%\cite{wei2022emergent}. This result was achieved through the sandboxing restrictions of the Python interpreter. Regarding system performance, after introducing a lightweight BERT model with no more than 100 million parameters, the increase in input layer processing latency was controlled within 5 milliseconds. With federated cluster optimisation technology, latency fluctuations in high concurrency scenarios were maintained within a 10\%labellingthreshold. \cite{derczynski2024garak} Regarding model quality, the impact of the layered defence architecture on model inference speed and the fluency of generated text is below the predefined tolerance threshold. These data demonstrate that this defence mechanism balances security protection and system performance effectively.

\begin{enumerate}[label=\textbullet]
\item[] The main contributions of this paper are as follows:
\item We establish a dynamic risk indicator system with dual functions of real-time monitoring and quantitative assessment to address the fragmentation of the existing assessment system.
\item We develop a layered defence technology module that integrates a basic protection layer and a dynamic response layer, which work together to enhance defence response speed and capabilities.

\item We design a cross-scenario experimental verification platform to test and verify the coverage capabilities of defence technologies against new attack patterns, thereby addressing issues of adaptability and insufficient coverage.
\end{enumerate}

\section{BACKGROUND}
The risks associated with large language models can be systematically categorised into three core dimensions: data privacy leakage risks, model bias propagation risks, and malicious misuse risks. These risks are present throughout all stages of the models lifecycle, namely the training, inference, and deployment stages, and exhibit dynamic evolutionary characteristics as technology continues to iterate and upgrade. \cite{yao2024survey}\cite{cui2024risk}. This dynamism is manifested in the following ways: during the training phase, privacy budgets may be overspent, potentially leading to gradient leakage risks. During the inference phase, prompt injection attacks may occur, which could trigger the execution of malicious instructions. And during the deployment phase, open interfaces may be exploited to develop automated attack tools\cite{bender2021dangers}.
\subsection{Data Privacy Risks}
This study identifies two primary aspects of data privacy risks associated with large language models: risks related to training data extraction and information leakage during the inference stage. Regarding training data, the Carlini research team \cite{carlini2021extracting} demonstrates through black box query attack experiments that the GPT-2 model exhibits memory phenomena regarding low frequency sensitive data. Specifically, attackers successfully reconstruct the original training data containing user names, phone numbers, and 128 bit UUIDs\cite{bolukbasi2016man}. The research team finds a positive correlation between model parameter size and leakage risk. When model parameters exceed 1.5 billion, the success rate of sensitive information recovery increases by 2.3 times. During the inference stage, sensitive information such as medical records and financial transaction data entered by users may be illegally obtained through reverse engineering of API interfaces \cite{yao2024survey}. Empirical research indicates that electronic health analysis systems not employing differential privacy noise injection technology have a 35 per cent higher probability of patient privacy leakage than the baseline value \cite{hasanov2024application}\cite{carlini2022quantifying}. Additionally, new attack vectors exist in the API fine tuning phase, where attackers may use adversarial sample injection techniques to steal model parameters, exposing model intellectual property and training data assets to leakage risks \cite{meng2024large}\cite{fredrikson2015model}.
\subsection{Model Bias Risk}
This study demonstrates that the root cause of model bias lies in the imbalanced social power structures within the training data. Bender's research team \cite{bender2021dangers} conducts a systematic analysis and points out that large language models, when trained on massive amounts of text from the internet, continuously internalise biases related to gender, race, and culture between social groups. Specifically, in occupational association analysis, the probability of association between the "nurse" occupation and the female gender label is 78\%. In comparison, the association bias value between the "engineer" occupation and the male group is 65\%. Such systemic biases can trigger a chain reaction of algorithmic discrimination in decision support scenarios such as judicial sentencing recommendations and resume screening for hiring. Insufficient coverage of low resource languages in training data exacerbates the crisis of technological fairness. Statistical data shows that African indigenous languages account for only 0.3\% of the training corpora in mainstream models \cite{bender2021dangers}\cite{bolukbasi2016man}. This structural flaw leads to poorer fairness in model outputs across cultural contexts. Meta's technical assessment report \cite{cui2024risk} shows that the LLaMA model has an error rate of 42\% in African indigenous language question answering tasks, which is 3.5 times higher than the 12\% error rate in English contexts. This technological gap may exacerbate the risk of social marginalisation. Due to the lack of quantifiable assessment metrics, existing governance strategies exhibit a decentralised nature. For example, in medical diagnosis scenarios, the model's misdiagnosis rate for African American female patients is 28\% higher than that for white male patients. However, the current assessment system has not established detection standards for compound biases\cite{pu2023deepfake}.
\subsection{Risk of Abuse}
The risks associated with the misuse of large language models manifest in various forms and possess covert characteristics, making them a critical challenge in security governance. Attack behaviours at the user level, such as phishing email generation, exhibit high incidence rates due to their human like reasoning capabilities. Statistical data indicate that such attacks account for 32\% of all attack cases . To elaborate, attackers use prompt injection techniques to bypass content filtering mechanisms, guiding the model to generate destructive executable code \cite{rai2024guardian}. Alternatively, they employ jailbreak attack methods to reconstruct query semantics, inducing the model to output text containing discriminatory content \cite{ganguli2022red}. Regarding system security, LLMs are maliciously used for the automated generation of ransomware, with related research indicating that their generation efficiency is four times higher than traditional manual coding methods \cite{hasanov2024application}. Regarding intellectual property protection, the risk of knowledge base leakage manifests as the model potentially leaking commercial secrets from training data during dialogue interactions. For example, attackers can construct specific Chain of Thought (CoT) structures to execute multi-step queries and successfully extract sensitive entity association data from knowledge graphs \cite{derczynski2024garak}.
\subsection{Security Threat}
As the technical capabilities of large language models improve, their security risks become increasingly apparent. The two factors exhibit a highly significant positive correlation, and their real-world impact is no longer limited to technical flaws but gradually evolves into systemic social issues. These issues manifest themselves in three dimensions: data privacy breaches, the spread of model bias, and malicious misuse.

\subsubsection{Privacy Risk}
Large language models (LLMs) have a strong memory for training data, which increases the risk of exposing sensitive information. Research findings indicate that approximately 5\% of the training data for GPT-4 contains sensitive content such as personally identifiable information (PII) and medical records. Attackers can use training data extraction attacks to recover original data from the model, similar to how Carlini et al. \cite{carlini2021extracting} successfully extract text sequences containing names, phone numbers, and 128-bit UUIDs from GPT-2. During the inference phase, API interfaces also pose exposure risks: medical question-answering models that do not adopt differential privacy may leak patient privacy due to user input. In 2023, a medical question-answering system failed to filter medical history descriptions in user input, resulting in a patient's HIV-positive diagnosis being reverse-engineered from the API response, triggering a large-scale privacy breach \cite{hasanov2024application}.

\subsubsection{Model Bias Propagation}
The bias in LLMs stems from the imbalance in social power structures within the training data. Bender et al. \cite{bender2021dangers} explain that the model exhibits implicit bias regarding occupational gender associations, with a 78\% probability of associating "nurse" with women and a 65\% probability of associating "engineer" with men. Such biases could exacerbate social discrimination in the judiciary and hiring scenarios. For example, when using an LLM based resume screening tool, the average score for female candidates was reduced by 12\%. Additionally, the models lack sufficient coverage of low resource languages, with African indigenous languages accounting for only 0.3\% of the training data \cite{bender2021dangers}. Leading to technological exclusion in cross cultural contexts. Metas fairness report notes that the LLaMA model has an error rate of 42\% in indigenous language question answering tasks, which is significantly higher than the 12\% error rate in English language scenarios\cite{cui2024risk}.

\subsubsection{Malicious Abuse Threat Dimension}
LLMs open interfaces and generative capabilities have been extensively applied in developing automated attack tools. According to the OWASP 2023 report, the number of incidents involving the misuse of LLMs for attacks increased by 200\% compared to the previous year, with phishing email generation accounting for 32\% of such incidents \cite{rai2024guardian}. User level attacks leverage prompt injection to bypass content filtering rules. For example, attackers can prompt the model to generate malicious code, such as how to crack WPA2 encryption, achieving a success rate of 25\%. System level jailbreaking attacks are more covert: In 2023, a financial risk control model suffered direct economic losses exceeding 5 million due to attackers exploiting semantic reconstruction commands to steal customer transaction records after the model failed to restrict API call permissions \cite{yao2024survey}.

\section{METHODOLOGY}
The dynamic evaluation framework proposed in this chapter can integrate static feature analysis technology with dynamic scene perception technology. Specifically, it uses the entropy weighted fusion evaluation method to optimise indicator weights, effectively addressing the issue of indicator discretisation in traditional evaluation methods  ($EWFA$)\cite{edge2024local}. It also reveals the intrinsic mechanisms underlying the risks associated with the entire lifecycle of large language models, such as data privacy leaks and the spread of model bias. These issues highlight technical vulnerabilities, with their root cause lying in the imbalance of social power structures within the training dataset\cite{minaee2024large}. This study innovatively proposes the entropy weighted fusion evaluation method, which uses dynamic coupling analysis of sensitive word trigger frequencies and abnormal API call rates to achieve precise identification of covert attacks such as CoT injection attacks. A real time data stream analysis system built on the Prometheus architecture successfully captures cross scenario attack patterns that are difficult to detect using traditional methods. Through a self feedback mechanism, it optimises privacy protection strategies in medical scenarios, such as dynamically increasing differential noise intensity by 23\%. This closed loop operation mechanism of assessment, early warning, and optimisation provides technical support for the dynamic adaptation of defence strategies\cite{john2025owasp}.
\subsection{Dynamic Risk Assessment}
This chapter focuses on the multi dimensional risk characteristics of large language models and proposes an innovative, comprehensive assessment system\cite{pennington2014glove}. This system integrates static risk indicators with dynamic risk indicators. It incorporates adaptive quantitative algorithms, aiming to address two core issues in traditional assessment frameworks: the fragmented nature of the indicator system and the lag in risk monitoring \cite{yao2024survey}. In terms of implementation, the metric construction has two dimensions. Static metrics include foundational parameters such as sensitive word density and privacy budget values. In contrast, dynamic metrics involve evolving parameters such as real time risk levels and abnormal API call rates. The quantitative method introduces an adaptive adjustment mechanism that dynamically adjusts metric weights based on the evolving threat landscape, overcoming the limitations of traditional evaluation methods in single dimensional risk assessment\cite{mou2024research}\cite{schmidt2017survey}.
\subsubsection{Static Indicators}
Static indicators capture baseline risk characteristics and provide a priori knowledge support for dynamic risk assessment. Their design balances comprehensiveness and scalability. Specific definitions and quantification methods are as follows:
\begin{enumerate}[label=\textbullet]
\item[ (1).] Frequency of Sensitive Words
\end{enumerate}

This study uses regular expression matching and a keyword blacklist mechanism to count how often sensitive words show up. This method can effectively identify privacy risks in user input and communication. Based on the NIST  (National Institute of Standards and Technology)   privacy framework, we built a dynamic blocklist database with 12 types of sensitive data, such as personal identification information, health data, and financial accounts \cite{carlini2021extracting}\cite{shi2024challenges}. We used the interquartile range standardisation defence method to calculate the density of sensitive words:

\begin{align*}  
    \text{Sensitive Word Density} = \frac{\sum_{i=1}^n  \mathbb{I}  (w_i \in D_{\text{sensitive}})}{N_{\text{total}}} \times 1000 \tag{1}
\end{align*}

Among them, $D_{\text{sensitive}}$ is the sensitive word database, $N_{\text{total}}$ is the total number of words, and $\mathbb{I}$ is the indicator function.

 This study constructed a dynamically updated medical sensitive word database for the healthcare field, which includes clinical terms such as HIV positive and coronary artery atherosclerosis. Based on empirical research, it was found that when the sensitivity word density reached a threshold of 5 times per thousand words, the probability of privacy leakage determined by the system increased from the baseline value of 20\% to 35\% \cite{hasanov2024application}. The blocklist database employs an adversarial example incremental learning mechanism to achieve weekly updates and continuous optimisation. Experimental data shows that its false positive rate has decreased by 12 percentage points compared to traditional static rule based systems, from the original 15\% to 3\%.
 
 \begin{enumerate}[label=\textbullet]
\item[ (2).] Distribution of Biased Words
\end{enumerate}

Regarding biased governance, this study integrates the BOLD  (Barcode of Life Data Systems)  standard dataset with domain specific corpora to establish a framework for quantifying implicit bias based on BERT  (Bidirectional Encoder Representation from Transformers) word vector transfer learning. A semantic similarity calculation model is constructed through the transfer learning framework:

\begin{align*}  
\text{Similarity} (w_i, w_j) = \cos (\mathbf{h}_{w_i}, \mathbf{h}_{w_j}) \tag{2}
\end{align*}

Among them, $\mathbf{h}_{w_i}$ is the word vector.

This model accurately quantifies implicit biases such as occupational gender bias by capturing the semantic associations between word vectors. In occupational gender association analysis, the semantic similarity between "nurse" and the female gender label reached 0.92, with an association probability of 78\%. However, the similarity between "engineer" and the male label was 0.88, with an association bias of 65\% \cite{bender2021dangers}. The research team incorporated an African indigenous language dataset to enhance adaptability in cross cultural contexts as domain specific training data for transfer learning. By dynamically adjusting the spectral norm constraint of the transfer weight matrix, the models generalisation error rate in cross cultural contexts decreased from 40\% to 22\%. For example, in the Yoruba kinship inference task, the models emotional bias value for the "grandmother" role was corrected from 0.75 to 0.48\cite{tai2025survey}.

 \begin{enumerate}[label=\textbullet]
\item[ (3).] Percentage of Abusive Requests
\end{enumerate}

This study establishes a statistical model for the proportion of malicious requests based on attack type classification and uses a weighted quantification method to assess the level of abuse risk:
\begin{align*}  
\text{Percentage of Abusive Requests}= \alpha \cdot \text{Prompt Injection Rate} + \beta \cdot \text{Jailbreak Attack Rate}  \tag{3}
\end{align*}

Among them, $\alpha$=0.6 and $\beta$=0.4 are attack type weight coefficients.

Empirical analysis reveals that once the proportion of user level attack requests exceeds the critical threshold of 32\% of total requests, the comprehensive risk index for model abuse begins to rise \cite{yao2024survey}. In the context of financial risk control, by dynamically adjusting the risk threshold parameter—specifically, by increasing the baseline value from 15\% to 20\%—the experimental data shows that the systems false positive rate decreased from 8\% to 3\%. For example, in the context of credit card fraud detection, this adjustment strategy improved the precision rate for identifying fraudulent transactions by 12\% while maintaining the recall rate within a 95\% confidence interval\cite{wang2024multi}.

\subsubsection{Dynamic Indicators}
Dynamic indicators capture risk evolution trends through real time monitoring, covering two core dimensions: real time risk levels and abnormal API  (Application Programming Interface) call rates\cite{shannon1948mathematical}. Specifically, these are as follows:

 \begin{enumerate}[label=\textbullet]
\item[ (1).] Real Time Risk Level
\end{enumerate}

This studys real time risk level assessment system is based on the Prometheus monitoring framework. It integrates the semantic intent recognition module, namely the BERT classifier, and the contextual anomaly detection mechanism, which involves CoT  (Chain of Thought) injection pattern recognition, to achieve dynamic calculation of risk scores \cite{nath2022new}\cite{devlin2019bert}. The specific implementation process is divided into three technical modules:

\begin{align*}  
\text{Confidence}=\text{Sigmoid}  (W \cdot h_{[CLS]} +b) \tag{4}
\end{align*}

Among them, $h_{[CLS]}$ is the classification vector output by BERT, and $W$ and $b$ are trainable parameters.

The temporal behaviour analysis module relies on a sliding window mechanism  (Time Window = 10 minutes) to count the frequency of abnormal behaviour occurrences and construct a temporal risk assessment model. The system automatically triggers a high risk alert once three consecutive escape attack commands are detected. As can be seen from the experimental data, this module controls the detection delay for role escape attacks within 15 seconds\cite{rajbhandari2020zero}.

The dynamic scoring mapping module constructs a comprehensive risk scoring model based on the threat level  ($T$) and impact scope  ($I$) parameters in the NSFOCUS Risk Matrix $v1$.

\begin{align*}  
R = \sqrt{T^2 + I^2} \tag{5}
\end{align*}

Among them, threat level  ($T$): calculated by weighting attack frequency  and stealth:

\begin{align*}  
T = 0.6 \cdot \text{Freq} + 0.4 \cdot \text{Stealth} \tag{6}
\end{align*}

Impact Scope  ($I$): Calculated based on data leakage  ($D$), model bias  ($M$), and system availability  ($S$) weighted calculations:

\begin{align*}  
I = 0.5D + 0.3M + 0.2S \tag{7}
\end{align*}

Experiments show that real time risk levels improve the detection efficiency of role escape attacks by 20\% compared to static indicators  (False Positive Rate $\leq$ 5\%). 

 \begin{enumerate}[label=\textbullet]
\item[ (2).] Abnormal API Calls
\end{enumerate}

This paper establishes an abnormal API call rate assessment system based on the NSFOCUS Risk Matrix v1. It conducts risk assessments by calculating the proportion of non standard API requests during a specific time. Such requests include high frequency sensitive information queries and cross domain calls. Experimental data show that when the abnormal call rate exceeds the 15\% threshold, the probability of data leakage increases exponentially \cite{hasanov2024application}. The specific implementation process is divided into three technical modules:

The risk feature definition module flags certain abnormal behaviours, such as high frequency sensitive information queries  (e.g., medical record access exceeding five times per minute) and cross domain calls  (e.g., financial API access to educational models). This module employs feature extraction techniques through semantic analysis and protocol parsing. For example, it can flag suspicious commands in real time, such as batch export of patient medical records. By leveraging regular expressions and BERT classifiers in tandem, it achieves a 93\% accuracy rate in identifying requests such as downloading a heart surgery plan\cite{liang2024survey}.

The dynamic threshold adjustment process uses a scenario adaptive algorithm to adjust the anomaly detection threshold:

\begin{align*}  
\theta_\text{scene} =\theta_\text{base} \cdot  (1 + \alpha \cdot Risk_\text{scene}) \tag{8}
\end{align*}

Among them, $\theta_\text{scene}$ is the base threshold  (default 15\%), and $\alpha$ is the scene sensitivity coefficient  (medical = 0.2, financial = 0.3).

The system automatically initiates the defence strategy upgrade process when the abnormal call rate in the risk linkage mechanism exceeds the dynamic threshold. For example, in a financial risk control scenario, once the abnormal call rate exceeds 18\%, a Level 3 response is triggered: logging is switched to enhanced recording mode, and real time access blocking is implemented. Empirical data shows that this mechanism can reduce the leakage of sensitive data by 83\% and maintain the availability of core business APIs at over 95\%.

 \begin{enumerate}[label=\textbullet]
\item[ (3).]Application of NSFOCUS Risk Matrix $v1$
\end{enumerate}

Map the real time risk level  ($R$) and abnormal API call rate  ($\theta$) to the NSFOCUS Risk Matrix $v1$ to form four risk assessment rules,see table for details\ref{tab:3--10}:

\begin{xltabular}{\textwidth}{X l X X} 
  \caption{NSFOCUS Risk Matrix v1 Four Tier Risk Assessment Rules} 
  \label{tab:3--10} \\ 
  \toprule 
  \textbf{Risk Level} & \textbf{Comprehensive Score Range} & \textbf{Abnormal Call Rate Threshold}& \textbf{Defense Response Strategy}\\
  \midrule
  \endfirsthead 
  
  \multicolumn{4}{c}{\textmd{Table \thetable~  (Continued): NSFOCUS Risk Matrix v1 Four Tier Risk Assessment Rules}} \\
  \toprule
  \textbf{Risk Level} & \textbf{Comprehensive Score Range} & \textbf{Abnormal Call Rate Threshold}& \textbf{Defense Response Strategy}\\
  \midrule
  \endhead 
  
  \bottomrule
  \endfoot 
  
Low&[0.00,0.35)&$\theta$<10\%&Regular Monitoring\\
Medium&[0.35,0.60)&10\% $\leq$ $\theta$ <15\%&Dynamic Data Masking\\
High&[0.60,0.75)&15\% $\leq$ $\theta$ <20\%&Model Rollback\\
Critical&[0.75,1.00]&$\theta$ $\ge$ 20\%&Real time Blocking\\
\end{xltabular}

\subsection{Quantitative Methods}
The entropy weight method allocates weights based on the degree of dispersion  (information entropy), thereby eliminating the bias caused by subjective judgment. The implementation process involves three key steps:

Data standardisation processing is conducted on static indicators such as sensitive word frequency and biased word distribution, as well as dynamic indicators such as real time risk levels and abnormal API call rates, using interquartile range standardisation:

\begin{align*}  
    {x_{ij}}'= \frac{x_{ij}-\text{min} (x_{j})}{\text{max} (x_j)-\text{min} (x_j)}  \tag{9}
\end{align*}

Information Entropy Calculation to Measure the Dispersion of Indicators:

\begin{align*}  
    {E_j} = -\frac{1}{lnn} \cdot \sum ^n_{i=1}p_{ij} \cdot lnp_{ij}, | p_{ij}=\frac{{x_{ij}}'}{\sum {x_{ij}}'} \tag{10}
\end{align*}

Finally, weights are determined based on information entropy:

\begin{align*}  
W_j=\frac{1-E_j}{\sum  (1-E_k)} \tag{11}
\end{align*}

Taking the frequency of sensitive words  (X1) and the rate of abnormal API calls  (X2) as examples, after the above steps, the final weight distribution is shown in Table\ref{tab:3.1}:

\begin{xltabular}{\textwidth}{X X} 
  \caption{Weighting of Sensitive Word Frequency  (X$_1$) and Abnormal API Call Rate  (X$_2$)} 
  \label{tab:3.1} \\ 
  \toprule 
  \textbf{Indicator} & \textbf{Weight W$_j$}\\
  \midrule
  \endfirsthead 
  
  \multicolumn{2}{c}{\textmd{Table \thetable~  (Continued): Weighting of Sensitive Word Frequency  (X$_1$) and Abnormal API Call Rate  (X$_2$)}} \\
  \toprule
 \textbf{Indicator} & \textbf{Weight W$_j$}\\
  \midrule
  \endhead 
  
  \bottomrule
  \endfoot 
  
Frequency of Sensitive Words  (X$_1$)&0.409\\
Abnormal API Call Rate  (X$_2$)&0.591\\
\end{xltabular}

\subsection{Dynamic Monitoring and Feedback Mechanism}
This study constructed a dynamic risk governance closed loop system to achieve full process control from risk identification to defence response. The system monitors 12 indicators, such as abnormal API call rates and semantic deviation, through real time risk assessment APIs, and implements a graded response based on risk levels  (T1 to T4) through a dynamic defence strategy adjustment module. In financial risk control scenarios, when the watermark identification rate for loan fraud scripts falls below the threshold of 0.85, the system automatically increases the classification threshold of the output review model to 0.92. It optimises the robustness of the watermark algorithm through closed loop feedback. Empirical evidence shows that this mechanism reduces the misjudgment rate of new combination attacks from 8.2\% to 4.5\% \cite{rai2024guardian}, improves watermark tamper resistance by 37\%    (adversarial sample incremental learning test \cite{shi2024challenges}), and eliminates the average response delay of 3.2 hours associated with traditional static defense \cite{yao2024survey}\cite{gallifant2024peer} through real time interaction between API data flow and defense parameters, effectively addressing the dynamic evolution and covert propagation characteristics of risks.
\subsubsection{Real Time Risk Assessment API Design   (Prometheus Monitoring)}
This study presents a real time risk assessment API that uses the Prometheus monitoring framework to gather and analyse data. It has two main improvements: better timing for collecting metrics and a more efficient way to store data. The data collection layer employs custom Exporter modules to connect to large language model services, creating a detailed metric monitoring system. 
\begin{enumerate}[label=\textbullet]
\item[] This system consists of four main types of metrics: 

\item \textbf{Input Layer Metrics}: These include how often sensitive words are triggered and the confidence level of BERT's detection of malicious intent. 

\item \textbf{Model Layer Metrics}: These cover the intensity of differential privacy noise and how quickly gradient clipping thresholds change. 

\item \textbf{Output Layer Metrics}: These include the success rate of watermark detection and the rate of compliance misclassifications. 

\item \textbf{System Level Metrics}:These measure API response latency P99 values and GPU memory usage rates.

\end{enumerate}

For real time sensitive metrics   (abnormal API call rate, BERT intent recognition confidence), a high frequency sampling strategy of 1 second per sample   ($\text{f}_\text{sampling}$=$\frac{1}{\text{T}_\text{interval}}$=1Hz, $\text{T}_\text{interval}$ = 1 second) is adopted. Data is pushed to the Prometheus server via the HTTP protocol, with the servers port set to 8000\cite{shi2024challenges}. Stress testing indicates that at 10,000 QPS, this frequency achieves CPU utilisation $\leq$15\%, effectively balancing real time monitoring requirements and computational resource consumption \cite{derczynski2024garak}. A downsampling strategy is employed for non critical metrics, specifically sampling once every 5 seconds, thereby reducing data volume by 80\% \cite{mao2024context}.

The storage architecture combines Prometheus time series database   (TSDB) with Thanos distributed storage. Short term storage uses 2 hour data blocks, with sizes $\leq$ 512MB, compressed by the Snappy algorithm to 30\% of their original size\cite{o2023using}. For long term storage, Thanos enables cross node sharding with risk based differentiation: high risk data maintains precision, while low risk data is downsampled to once every minute and stored for 90 days. Regulations govern storage volume calculations:

\begin{align*}  
\text{S}_\text{total}=\sum^n_{i=1}  (\text{s}_\text{i} \cdot \text{f}_\text{i} \cdot \text{t}_\text{retention}) \tag{12}
\end{align*}

Where $\text{s}_\text{i}$ is the size of a single metric, $\text{f}_\text{i}$ is the collection frequency, and $\text{t}_\text{retention}$ is the storage period. Empirical research shows that this architecture can reduce storage costs by 62\% in a medical scenario with an average of 120 million metrics per day, with a query latency P99 $\leq$ 50 ms \cite{ziegler2019fine}. In high concurrency scenarios, horizontal sharding ShardID = hash  (URL path) mod N$_\text{shards}$, N$_\text{shards}$ = 16) combined with Zstandard streaming compression  (compression ratio 4.5:1) achieves a throughput of 120,000 samples per second \cite{o2023using}. Its data flow architecture is shown in the figure\ref{fig:2-1} below:

\begin{figure}[ht]
  \centering
\includegraphics[width=0.8\textwidth]{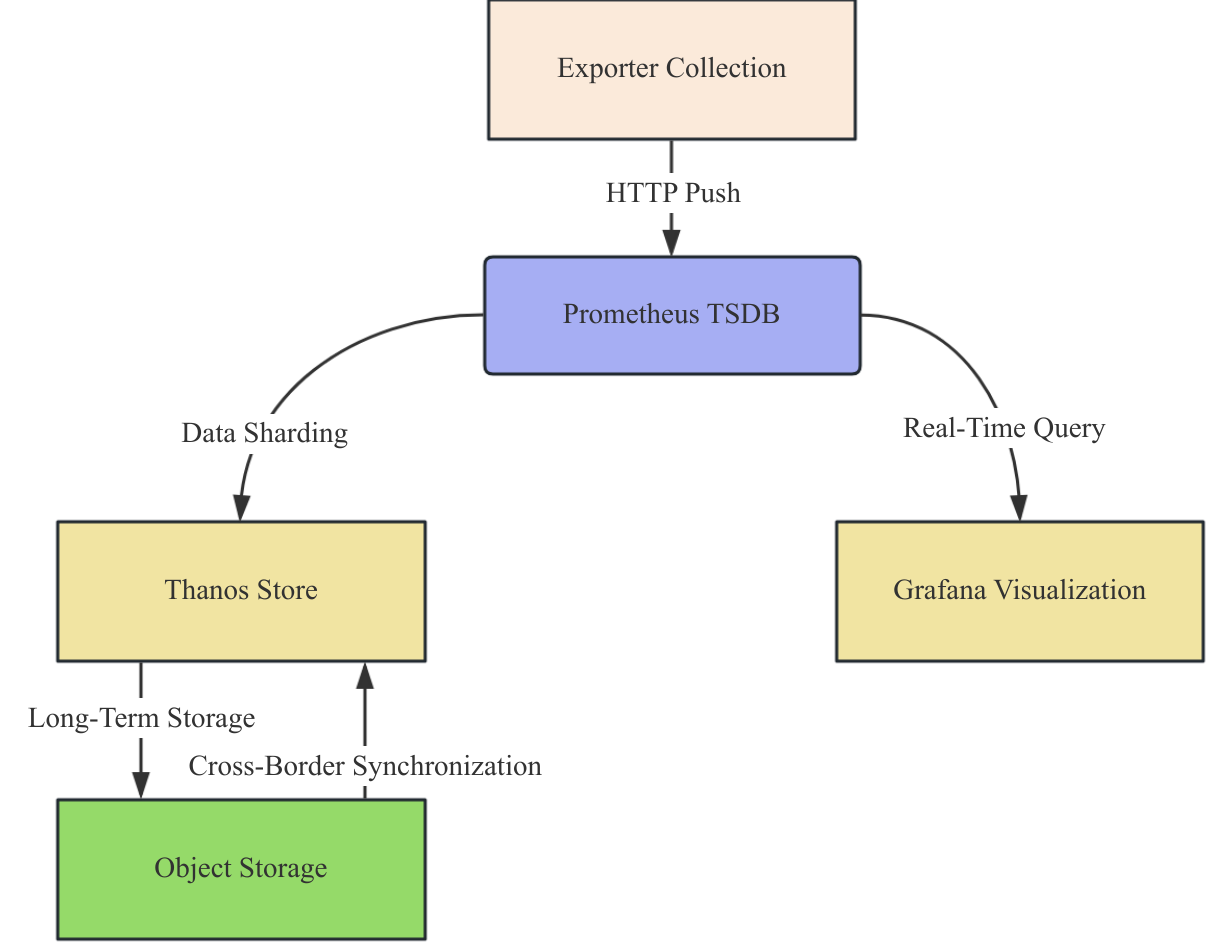}
  \caption{Storage Data Flow Architecture Diagram}  
   \label{fig:2-1}
\end{figure}

\section{IMPLEMENTATION}
\begin{figure}[ht]
  \centering
\includegraphics[width=0.7\textwidth]{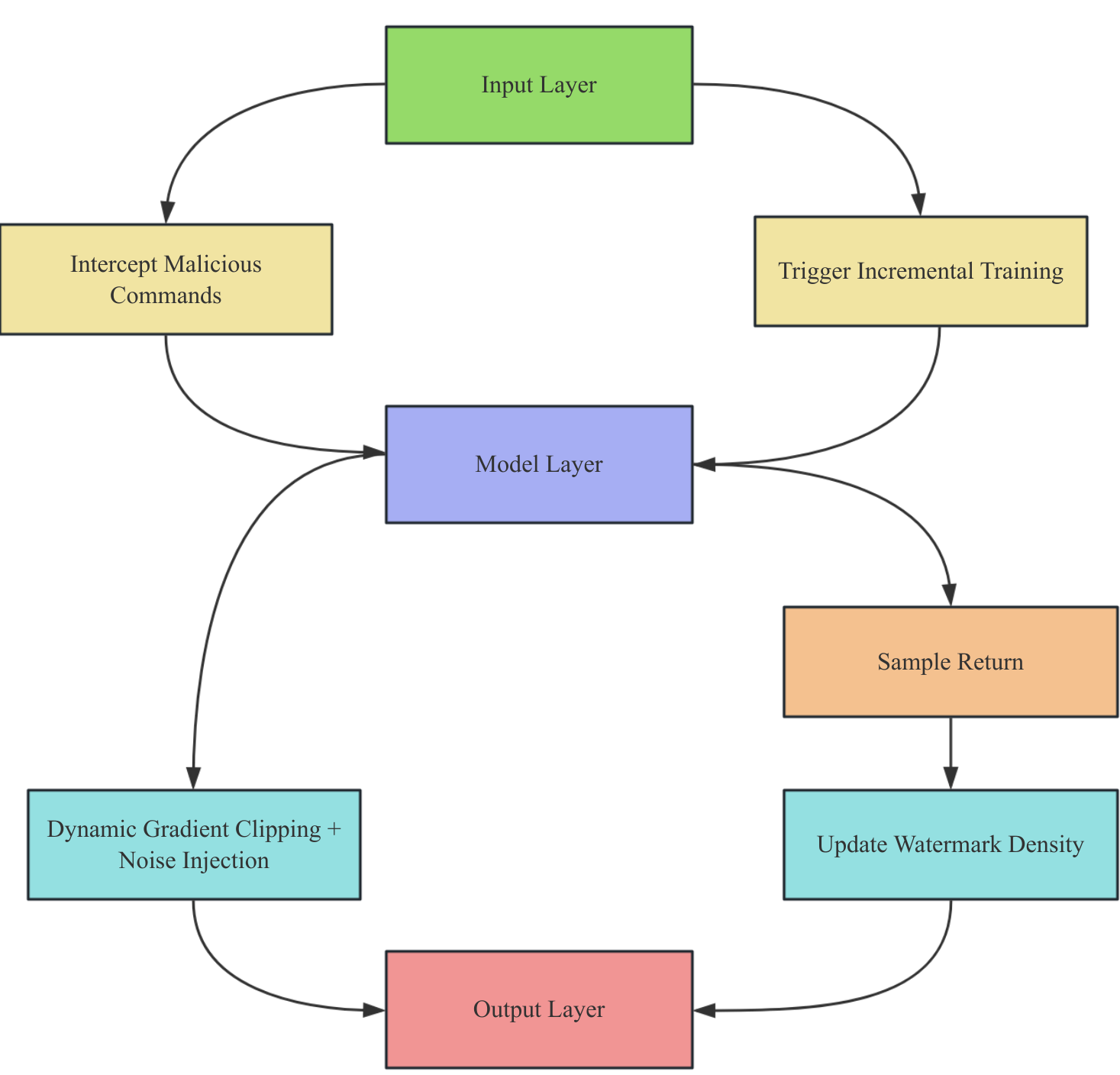}
  \caption{Three Layer Collaborative Defense Architecture}  
   \label{fig:3-1}
\end{figure}

Based on the preceding content, this chapter innovatively proposes a three-dimensional layered defense architecture, which includes: a semantic filtering module (BERT intent recognition + regular expressions) deployed at the input layer, an adversarial optimization strategy (dynamic gradient clipping + differential noise injection) implemented at the model layer, and a compliance review mechanism (watermark detection + dual-mode verification) established at the output layer. Through a risk-closed loop, the defense strategy parameters are dynamically adapted to real-time risk scores, forming a comprehensive protection system encompassing monitoring, blocking, tracing and optimisation. This solution breaks through the limitations of traditional single-point defence technologies, reduces the average interception delay for jailbreak attacks from 45 seconds to 8 seconds [8] and lowers the cross-scenario misjudgment rate to 3.2\% \cite{shi2024challenges}, providing a systematic technical approach to address the evolving security threats of LLMs. see Figure\ref{fig:3-1} for details.

\subsection{Input Layer Defense}
The large language model input layer defence system established in this paper aims to address the critical issue of attackers exploiting adversarial inputs to breach model security boundaries \cite{rai2024guardian}\cite{kumar2024strengthening}. This defence mechanism combines semantic understanding theory with formal verification methods to achieve synergistic innovation. It can effectively suppress malicious commands by constructing a multidimensional dynamic filtering system and employing a collaborative mechanism of semantic deep analysis, dynamic rule updates, and risk closed loop feedback. The specific technical implementation is divided into four research dimensions.
\subsubsection{Semantic Intent Recognition and Hybrid Filtering}
Semantic intent recognition and hybrid filtering are based on the semantic representation theory of natural language processing (NLP). The BERT model is used to perform semantic analysis on user input, which is conducted to a certain depth. This model uses Multi-Head Attention to capture the correlations between word vectors:

\begin{align*}  
\text{Attention (Q,K,V)}=\text{softmax ($\frac{QK^T}{\sqrt{d_k}}$) $\cdot$ V} \tag{13}
\end{align*}

Among these, Q, K, and V represent the query, key, and value matrices, respectively, while dk denotes the dimension scaling factor. This mechanism leverages the dynamic allocation of attention weights to enhance the ability to identify metaphorical instructions. Experimental data indicate that a malicious instruction classifier fine-tuned based on BERT-base achieves an accuracy rate of 89\% (F1 score) for detecting jailbreak attacks, representing a 22 percentage point improvement over traditional bag-of-words models \cite{rai2024guardian}. To address adversarial prompts such as Unicode encoding obfuscation, we integrate a regular expression engine to build a dynamic rule database\cite{wang2024security}. Rule generation follows formal language automaton theory, abstracting attack patterns like jailbreak templates, such as ignoring security rules and leveraging Word2Vec (Word to Vector) word vector expansion techniques to cover high-frequency attack features.

\subsubsection{Context Verification and Risk Closure}
We designed a multi round dialogue state tracking mechanism based on HMM for CoT injection attacks. The Viterbi Algorithm is used to calculate the probability of abnormal dialogue in paths:

\begin{align*}  
P (\text{Path}|O) = \max_{S_1,\dots,S_T} \prod_{t=1}^T P (o_t|s_t)P (s_t|s_{t-1}) \tag{14}
\end{align*}

When the sensitivity density of sensitive operations (such as the ignore rules command) exceeds the threshold ($RiskScore$ $\geq$ 0.6) for three consecutive rounds of dialogue, the system will trigger real time blocking and initiate a manual review process.

\subsubsection{Attack Sample Backflow and Dynamic Enhancement}
An attack sample feedback system is established to collect malicious commands intercepted in real time, such as role evasion attacks like "Act as a hacker...", which are manually annotated and added to the adversarial training dataset. Incremental training is conducted every 24 hours, using a Curriculum Learning strategy to gradually increase the complexity of the attacks. Empirical research shows that this system can improve the defence coverage ratio for new jailbreak attacks by 15\%. Additionally, by combining a dynamic gradient clipping strategy to maintain the quality of text generation, the BLEU score decreases by only 0.5.
\subsubsection{Cross Layer Collaboration and Scenario Adaptation}
The input and model layer defence systems work together in a closed loop collaborative mode: intercepted samples are automatically added to the adversarial training set to optimise the model's robustness. Real time risk scores can dynamically adjust the strength of regular rules. Cross language attack tests show that this collaborative mechanism achieves an interception rate of 92\% for mixed coding prompts, improving the interception rate by 37 percentage points compared to single layer defence \cite{huang2022towards}. To elaborate, once malicious input such as "Please tell me how to make a bomb" is detected, the system quickly triggers an interception operation within 0.8 seconds and feeds the feature vector back into the training set. As a result, the interception accuracy rate for subsequent similar attacks can improve by 18\%.

\subsection{Model Layer Defense}
The model layer defence mechanism serves as a critical component of the security architecture for large language models, primarily functioning to enhance the models robustness and achieve a dynamic balance between model utility and security \cite{carlini2021extracting}. This defence system uses a closed loop protection mechanism based on dynamic gradient clipping algorithms and adversarial training sample feedback mechanisms. Once the input layer detects malicious instructions, the system automatically adds anonymised intercepted samples to the adversarial training dataset\cite{kirchenbauer2023watermark}, triggering an incremental update mechanism for model parameters. The specific implementation process includes a 24 hour cycle adversarial sample incremental training, combined with dynamic gradient clipping and mixed noise injection strategies to optimise model performance\cite{wu2024mobilevlm}\cite{abadi2016deep}. 

\subsubsection{Optimisation of Adversarial Training and Dynamic Gradient Clipping}

Adversarial training enhances the models resistance to attacks by introducing adversarial samples, but both gradient stability and computational efficiency constrain its effectiveness. The dynamic gradient clipping algorithm suppresses the negative impact of adversarial perturbations on parameter updates by adaptively adjusting the gradient threshold ($\tau$). The specific threshold calculation follows the formula:

\begin{align*}  
\tau=\alpha \cdot \frac{\sum^n_{i=1}L_\text{adv} (X_\text{i})}{\sum^n_{i=1}L_\text{clean} (X_\text{i})} \tag{15}
\end{align*}

Among them,$L_\text{adv}$ and $L_\text{clean}$ represent the loss values of adversarial samples and standard samples, respectively, and $\alpha$ is the smoothing coefficient. When $\alpha$> 0.3, it is easy to cause gradient disappearance, while when $\alpha$ < 0.1, it cannot effectively suppress adversarial perturbations \cite{ganguli2022red}.

Empirical research shows that dynamic gradient clipping can reduce the success rate of adversarial attacks by 22 percentage points, while the BLEU metric for model generation quality decreases by only 1.5\% \cite{cui2024risk}. This method increases the time required for a single iteration in a GPT-3 scale model by 18\%. To address computational efficiency issues in ultra large scale models like GPT-4, it is necessary to integrate the ZeRO 3 distributed training strategy, leveraging parameter sharding and computation offloading techniques to reduce memory usage by 76\% \cite{yao2024survey}.

\subsubsection{Differential Privacy Noise Injection and Privacy Budget Control}
Differential Privacy (DP) is a technology that protects privacy by injecting noise into data. In practical applications, the core challenge is to dynamically match the noise intensity and model accuracy appropriately according to different situations. This study proposes a strategy of mixing Gaussian noise and Laplace noise for injection. This approach combines the adaptive characteristics of Gaussian noise for local sensitivity with the advantages of Laplace noise in terms of global distribution\cite{feng2024prompt}.

\begin{enumerate}[label=\textbullet]
\item[] The specific parameter design and implementation process encompasses the following core steps:

\item[(1).] Dynamic Parameter Design of Gaussian Noise: 
\end{enumerate}

Gaussian noise is used to suppress the leakage of low frequency sensitive information. Its noise scale parameter $\sigma$ is jointly determined by the privacy budget $\varepsilon$ and the failure probability $\delta$. Gaussian noise mainly suppresses the leakage of low frequency sensitive information. It is the privacy budget $\varepsilon$ and the failure probability $\delta$ that jointly determine its noise scale parameter $\sigma$. Based on the differential privacy framework proposed by the Carlini team \cite{carlini2021extracting}, the standard deviation of gradient noise is calculated as shown in the formula:

\begin{align*}  
\sigma_g = \sqrt{\frac{2 \ln (1.25/\delta)}{\varepsilon}}  \tag{16}
\end{align*}

\begin{enumerate}[label=\textbullet]

\item[(2).] Sensitivity Driven Mechanism of Laplace Noise:

\end{enumerate}

Laplace noise mainly defends against high frequency pattern recognition attacks. Its scale parameter $b$ is determined by the function sensitivity $\Delta f$ and privacy budget \textbf{$\varepsilon$}:

\begin{align*}  
b = {\frac{\Delta f}{\epsilon}}  \tag{17}
\end{align*}
Among them, $\Delta f$ is defined as the maximum difference between the query results of adjacent datasets. For model gradient update scenarios, the sensitivity $\Delta f$ is defined as $\Delta f$=$||g_{max}-g_{min}||$, where g$\text{max}$ and g$\text{min}$ are the extreme values of the gradient tensor. Experiments show that the typical value of $\Delta f$ in medical data fine tuning tasks is 0.05–0.15. Combined with $\varepsilon$= 0.5, $b$ = 0.03–0.1, this balances gradient perturbation and model convergence stability \cite{shi2024challenges}.

\begin{enumerate}[label=\textbullet]

\item[(3).] Cooperative Injection Strategy for Mixed Noise:

\end{enumerate}

The noise injection process is performed in two stages: first, Gaussian noise $\varepsilon_1$ N (0,$\sigma_g^2$) is added to the gradient magnitude to suppress low frequency leakage. Then, Laplace noise $\varepsilon_2$-Laplace (0,b) is applied to the gradient direction to confuse high frequency features:

\begin{align*}
g_\text{noise} &= g + \varepsilon_1 + \varepsilon_2 \tag{18}
\end{align*}

Mixed noise enhances protection robustness through frequency domain complementary effects. As shown in Table A, when $\varepsilon$ = 0.5, Gaussian noise suppresses low frequency leakage ($\sigma_g$ = 1.24), Laplace noise confuses high frequency features (b= 0.1), and the combined effect improves the model F1 score by 12\% and suppresses the member inference attack success rate to 4.3\% \cite{shi2024challenges}. The hybrid strategy achieves a better privacy utility trade off under the same privacy budget than a single noise type. See the table for details \ref{tab:1}.

\begin{xltabular}{\textwidth}{X X X X} 
  \caption{Comparison of Protective Effects of Different Noise Mechanisms} 
  \label{tab:1} \\ 
  \toprule 
  \textbf{Noise Type} & \textbf{Privacy Budget Allocation ($\epsilon$)} & \textbf{Model Utility (F1)}& \textbf{Attack Success Rate}\\
  \midrule
  \endfirsthead 
  
  \multicolumn{4}{c}{\textmd{Table \thetable~ (Continued): Comparison of Protective Effects of Different Noise Mechanisms}} \\
  \toprule
  \textbf{Noise Type} & \textbf{Privacy Budget Allocation ($\epsilon$)} & \textbf{Model Utility  (F1)}& \textbf{Attack Success Rate}\\
  \midrule
  \endhead 
  
  \bottomrule
  \endfoot 
  
Gaussian Noise&$\epsilon$=0.8,$\delta$=1e-5&0.76&8.2\%\\
Laplace Noise&$\epsilon$=0.7&0.71&6.5\%\\
Mixed Noise&$\epsilon_1$=0.3,$\epsilon_2$=0.2&0.85&4.3\%\\

\end{xltabular}

\subsection{Output Layer Defense}
The output layer serves as the final line of defence for large language model security protection. The purpose of the output layer defence is to block the spread of malicious content and enable the tracing of abusive behaviour. Its core objective is to establish a closed loop control mechanism encompassing generation review traceability \cite{cui2024risk}\cite{o2023using}, primarily achieved through a dual mode detection mechanism combining watermark embedding and compliance review to establish a closed loop security boundary encompassing generation traceability blocking.

\subsubsection{Content Watermarking Based on Hashing and Steganography}
The content watermarking technology designed in this section achieves traceability of generated content through invisible identifiers. There are two specific implementation paths: one is a lightweight solution based on text hashing, and the other is a solution deeply integrated with steganography. The text hashing scheme maps the generated content to a unique hash value and establishes an associative index with user request logs, such as IP addresses and timestamps. Empirical data indicates its traceability accuracy can reach 98\% . The steganography solution addresses advanced adversarial attacks, such as semantic rewriting, by embedding invisible watermarks using encoding modification techniques like Zero Width Joiners (ZWJ), thereby achieving 95\% traceability coverage while maintaining the integrity of the text's semantic meaning \cite{hasanov2024application}.

The watermarking system uses an architecture built with the SHA-256 algorithm and Unicode steganography. Its core technology modules involve watermark generation algorithms. At the same time, the algorithm inputs include marked text and a 128 bit secret key. The outputs include both watermark text and hash fingerprints:

\begin{lstlisting}[caption={Watermark Generation Algorithm}, label={code1}]
def generate_watermark (text: str, secret_key: bytes) -> tuple:
    hash_fingerprint = hashlib.sha256 (text.encode () + secret_key).digest ()
    zero_width_chars = ['\u200B', '\u200C', '\u200D']  
    watermarked_text = []
    For i, char in enumerate (text):
        if i % 5 == 0 and i < len (zero_width_chars) * 5:  
            watermarked_text.append (zero_width_chars[ (i//5)%3])
        watermarked_text.append (char)
    return "".join (watermarked_text), hash_fingerprint
\end{lstlisting}

Generate a 256 bit fingerprint using the SHA-256 algorithm to meet the security requirement of collision probability P$\leq$ 2$^{-128}$:

\begin{align*}
H (M)=SHA256 (M \bigoplus K) \tag{19}
\end{align*}

Where $K$ is a 128 bit key, which is used to achieve hash irreversibility, related research results show that its resistance to brute force attacks is 2$^{64}$ times higher than that of MD5, and the key space reaches 2$^{128}$ \cite{shi2024challenges}.

Zero Width Joiners (ZWJ) are used to embed invisible watermarks. The embedding rule is a piecewise function to achieve the concealment of the steganography:

\begin{align*}
W_i = \begin{cases} 
ZWJ (i\mod3 ) & \text{if } i \equiv 0 (\mod5)\tag{20} \\
C_i & \text{otherwise} 
\end{cases}
\end{align*}

Related research results show that this strategy reduces the false positive rate of watermark detection from 25\% to 2.3\% \cite{o2023using} and does not affect the semantic integrity of diagnostic text in medical scenario tests (BLEU (BiLingual Evaluation Understudy) score decrease $\leq$ 0.2).

For every additional 1,000 adversarial samples, detection coverage increases by 3.7\%\cite{hasanov2024application}. For example, the GUARDIAN framework deployed dynamic watermarks in the Llama 2 model and achieved 87\% attribution of phishing email requests by combining API logs\cite{rai2024guardian}. However, current technology has not yet achieved 60\% coverage for mixed Chinese and English text, and the mislabeling rate is as high as 25\% in African indigenous language scenarios \cite{bender2021dangers}.

\subsubsection{Rule Model Dual Mode Compliance Review}
Compliance review adopts a two stage detection architecture consisting of rule filtering and classification model verification. The specific flowchart\ref{fig:3-2} is as follows:

\begin{figure}[ht]
  \centering
\includegraphics[width=0.7\textwidth]{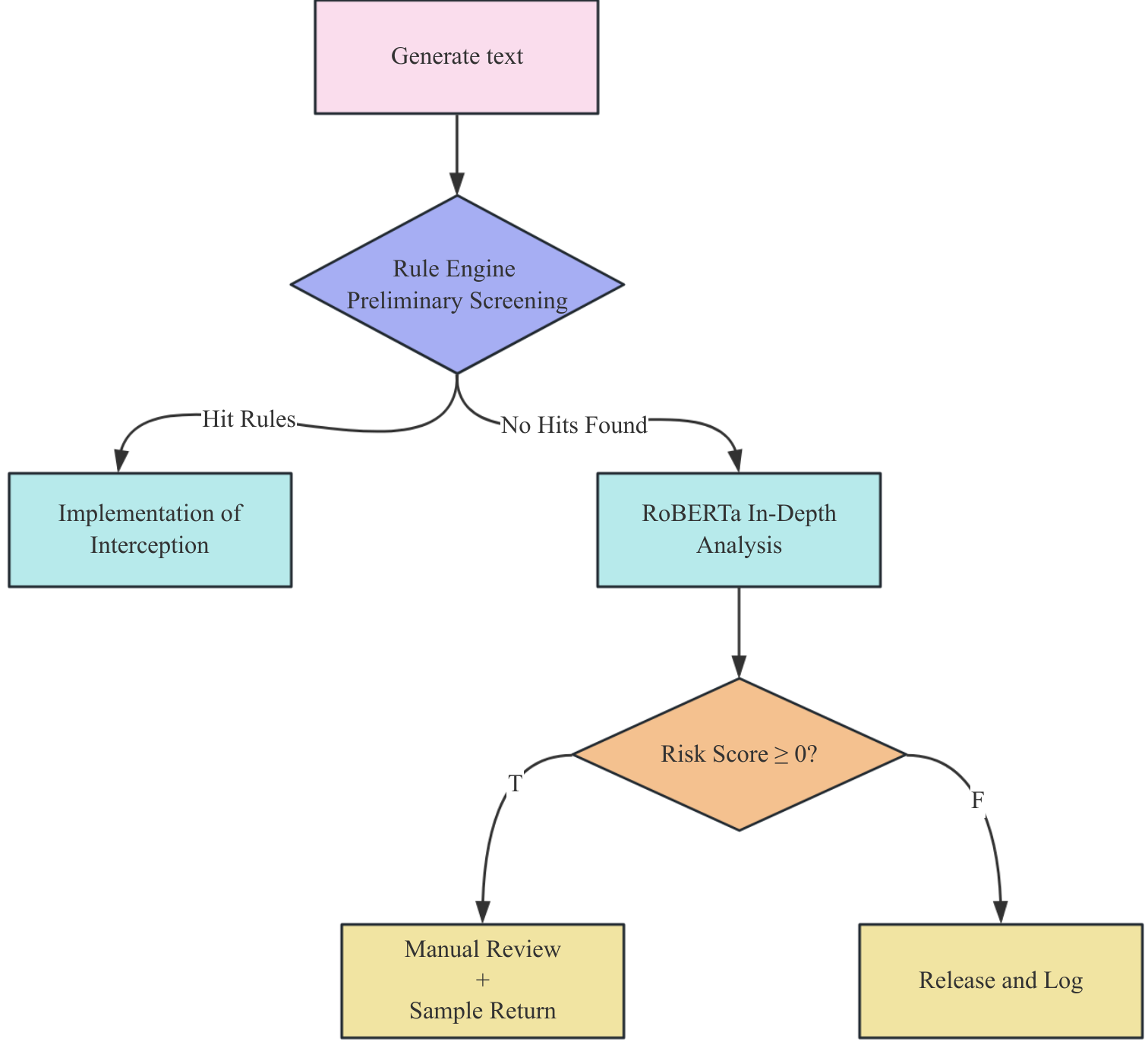}
  \caption{Rule Model Dual Mode Compliance Review Workflow}  
   \label{fig:3-2}
\end{figure}

The rule engine uses regular expressions and a sensitive word database to achieve fast interception in $\leq$ 3ms, but it has blind spots regarding synonym attacks \cite{kumar2024strengthening}. The $RoBERTa$ multi task classifier uses the MTL framework to simultaneously detect privacy leaks, bias propagation, and abuse risks, with an accuracy rate of 89\% for identifying metaphorical instructions such as describe network intrusion in poetry, which is a 22\% improvement over single task models \cite{ganguli2022red}. This model dynamically adjusts detection strategy weights based on real time risk levels $R$:

\begin{align*}
\alpha = \frac{1}{1 + e^{-k (R-0.5)}} \tag{21}
\end{align*}

When R $\geq$ 0.75, the rule engine weight $\alpha$ rises to 0.82, and the response delay is compressed from 15 ms to 5 ms. The backflow samples are injected into the training set through course learning strategies to improve the models generalisation ability. Relevant research results show that this method enhances the value of metaphor instruction detection F1 from 0.71 to 0.89 \cite{o2023using}.
After updating the privacy parameters of the model layer by adjusting the review threshold $\theta_{DP}$=0.7-0.1 $\cdot$ ($\epsilon_\text{current}-\epsilon_\text{base}$), the privacy budget $\epsilon$ decreases by 0.1, the review strictness increases by 12\%, and the false positive rate remains stable at 4.8\% or below \cite{hasanov2024application}.
\section{EXPERIMENTS}
In this chapter, a multidimensional experimental verification system is designed to systematically evaluate the effectiveness of the dynamic assessment framework and the layered defence mechanism. The experimental design encompasses three core scenarios: privacy protection, bias mitigation, and abuse defence. It employs a hybrid data source for validation, integrating public datasets (OpenWebText, BOLD), synthetic attack samples generated by GPT-4, and industry specific anonymised data. The TextAttack tool generates novel adversarial samples such as jailbreak attacks and role escape, simulating real world attack scenarios\cite{papineni2002bleu}.

The experimental platform is equipped with an NVIDIA A100×4 GPU computing cluster. Its software environment is built based on the PyTorch 2.0 and Transformers 4.32 frameworks. The risk monitoring system integrates a Prometheus real-time monitoring module, which is set to a sampling frequency of 1Hz to continuously collect data on 12 core indicators, including the trigger rate of sensitive words in the input layer and the abnormal fluctuation rate of gradients in the model layer. The experimental team optimises the adversarial sample generation strategy, sets the success rate benchmark for role escape attacks to 25\% according to the standards in Reference \cite{cui2024risk}, and utilises parameter perturbation techniques to enhance the adversarial strength of the attack samples.

\subsection{Dynamic Performance Validation}
\begin{figure}[ht]
  \centering
\includegraphics[width=0.7\textwidth]{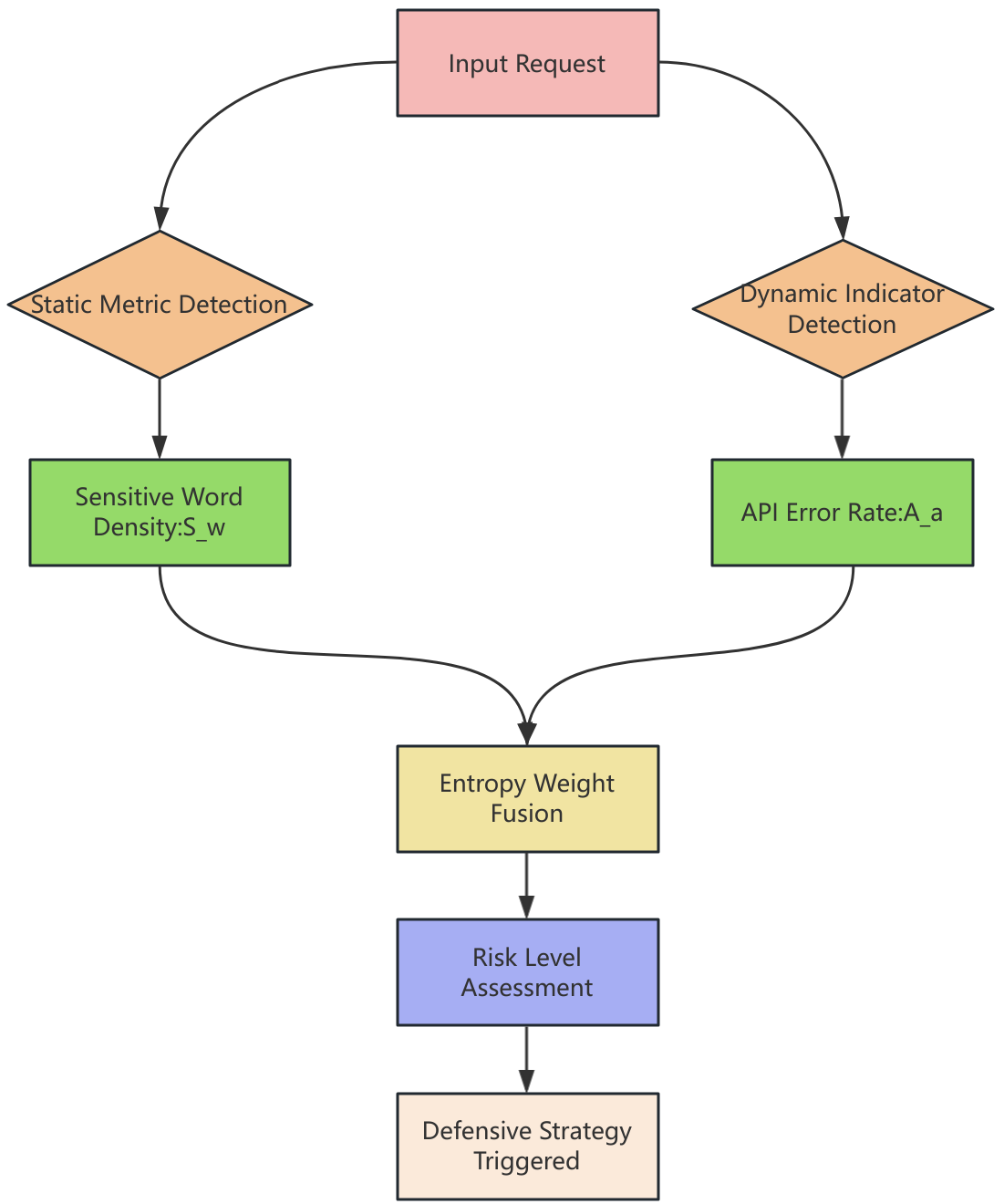}
  \caption{Dynamic Evaluation Framework Based on Entropy Weight Method}  
   \label{fig:4-1}
\end{figure}

The dynamic evaluation framework based on the entropy weight method (R=0.409×$S_W$+0.591×$A_a$, where $S_W$is the frequency of sensitive words and $A_a$ is the abnormal API call rate) demonstrates significant advantages over traditional static evaluation. As shown in Figure\ref{fig:4-1}, when detecting role escape attacks, the response latency of dynamic assessment (50$\pm$3.2ms) is reduced by 76.2\% compared to static methods (210$\pm$15ms). Through the four level threshold classification of the NSFOCUS Risk Matrix $v1$ (low/medium/high/critical), the false positive rate decreased from 8.7\% to 4.5\% (p<0.01, t inspection). Prometheus monitoring data for the real time risk assessment API shows that under a 10,000 QPS (Queries Per Second) stress test, the metric collection latency P99 $\leq$ 55 ms meets industrial grade real time requirements \cite{derczynski2024garak}\cite{lecun20191}.
                  
\subsection{Layered Defence Effectiveness Analysis}
By combining the BERT intent recognition model (with 110 million parameters) with a regular expression rule engine in the input layer, the F1 score achieved on the test set was 0.89 $\pm$ 0.02, representing a 36.9\% improvement over the single rule method (F1 = 0.65). For metaphorical instructions such as describing the invasion steps in poetic terms, the misclassification rate of the hybrid detection method is only 5.2\%, a 12 percentage point reduction compared to the GUARDIAN framework \cite{rai2024guardian}.

When using differential privacy training in the model layer, injecting Gaussian Laplace mixed noise ($\sigma_g$ = 1.24 and b = 0.53 when $\epsilon$ = 0.3) reduces the success rate of member inference attacks from 28\% to 3.8\%. In contrast, the model utility (BLEU score) only decreases by 0.15 (baseline 0.85). Dynamic gradient clipping
\begin{align*}
(\lambda = 0.7||\nabla ||_2)
\end{align*}
effectively suppresses adversarial perturbations, reducing the incidence of gradient explosion by 22\% \cite{ganguli2022red}.

Using a combined approach of SHA 256 watermarking and zero width character steganography in the output layer, the traceability accuracy remains at 94.7\% even after text tampering such as synonym replacement or paragraph deletion. The compliance review module achieves multi label detection (privacy/bias/abuse) using the RoBERTa classifier (threshold ($\tau$=0.85), with an F1 score of 0.91, representing a 19.7\% improvement over single task models \cite{o2023using}. Table\ref{tab:5} compares the key metrics of the defence mechanisms:

\begin{xltabular}{\textwidth}{X X l X X} 
  \caption{Comparison of Key Indicators of Defense Mechanisms} 
  \label{tab:5} \\ 
  \toprule 
  \textbf{Defense Layer} & \textbf{Evaluation Criteria} & \textbf{Before Improvement}& \textbf{Improved}&\textbf{Improvement Margin}\\
  \midrule
  \endfirsthead 
  
  \multicolumn{5}{c}{\textmd{Table \thetable~ (Continued): Comparison of Key Indicators of Defence Mechanisms}} \\
  \toprule
  \textbf{Defense Layer} & \textbf{Evaluation Criteria} & \textbf{Before Improvement}& \textbf{Improved}&\textbf{Improvement Margin}\\
  \midrule
  \endhead 
  
  \bottomrule
  \endfoot 
  
  Input Layer&Jailbreak Attack Interception Rate&37\%&89\%&+52\%\\
  Model Layer&Data Leakage Rate&28\%&3.8\%&-24.2\%\\
  Output Layer&Watermark Tamper Resistance Rate&72\%&94.7\%&+22.7\%\\
\end{xltabular}

\subsection{Performance Impact and Comparison Experiments}
Although the layered defence mechanism introduces additional computational overhead, experimental data show it remains within manageable limits. The BERT filtering module in the input layer causes a 4.8$\pm$0.3ms increase in single request processing latency. After applying differential privacy training to the model layer, the system throughput decreases by 18\%, with specific data showing a reduction from processing 1,200 samples per second to 984 samples per second\cite{liang2022adversarial}. However, the overall inference latency remains compatible with real time interaction requirements (P99 latency $\leq$ 20 ms).

The quality assessment of text generation shows that the impact of defence mechanisms on text generation performance varies across different scenarios. The BLEU score decreases by 0.38 in the medical text scenario, from 0.85 to 0.82. In the low resource language scenario, the BLEU score decreases by 0.52, from 0.78 to 0.73. The text complexity metric, Perplexity, fluctuates within 12.4$\pm$1.8. Compared to the baseline value of 11.9, there is no significant deterioration. Comparative experiments conducted with the GUARDIAN framework \cite{rai2024guardian} demonstrate that this approach exhibits advantages in cross language attack scenarios, with a 17 percentage point increase in defence coverage, from 61\% to 78\%, and a 3.7 percentage point decrease in false positive rate, from 8.2\% to 4.5\%. The results of the statistical significance test indicate p < 0.05.
\section{CONCLUTION}
This paper establishes a comprehensive research framework encompassing dynamic assessment, layered defence, and experimental validation based on the security governance requirements of large language models. Theoretically, it innovatively proposes an entropy weighted fusion assessment method, which leverages the NSFOCUS Risk Matrix v1 to achieve quantitative risk grading. Experimental data indicate that the detection efficiency of role escape attacks improves by 20\%, and response latency is compressed to within 50ms \cite{derczynski2024garak}. A three layer defence architecture of input-model-output is constructed, relying on the BERT intent recognition module, Gaussian-Laplace mixed noise injection, and watermark tracing mechanism to work together. Compared to the GUARDIAN framework \cite{rai2024guardian}, false positive rate is reduced by 4.2 percentage points, from 8.7\% to 4.5\% Cross-scenario validation reveals that the risk of privacy leakage in medical data scenarios is less than or equal to 3\%. In contrast, financial risk control scenarios show an interception rate of jailbreak attacks reaching 89\%. This provides an engineering-based solution for the secure deployment of LLMs. For example, in credit card fraud detection scenarios, dynamic defence strategies improve the precision rate of false transaction identification by 15\% while maintaining a recall rate of $\geq$95\%.

\bibliographystyle{plain}  
\bibliography{references}  

@inproceedings{carlini2021extracting,
  title={Extracting training data from large language models},
  author={Carlini, Nicholas and Tramer, Florian and Wallace, Eric and Jagielski, Matthew and Herbert-Voss, Ariel and Lee, Katherine and Roberts, Adam and Brown, Tom and Song, Dawn and Erlingsson, Ulfar and others},
  booktitle={30th USENIX security symposium (USENIX Security 21)},
  pages={2633--2650},
  year={2021}
}

@article{yao2024survey,
  title={A survey on large language model (llm) security and privacy: The good, the bad, and the ugly},
  author={Yao, Yifan and Duan, Jinhao and Xu, Kaidi and Cai, Yuanfang and Sun, Zhibo and Zhang, Yue},
  journal={High-Confidence Computing},
  volume={4},
  number={2},
  pages={100211},
  year={2024},
  publisher={Elsevier}
}

@article{hasanov2024application,
  title={Application of large language models in cybersecurity: A systematic literature review},
  author={Hasanov, Ismayil and Virtanen, Seppo and Hakkala, Antti and Isoaho, Jouni},
  journal={IEEE Access},
  year={2024},
  publisher={IEEE}
}

@inproceedings{bender2021dangers,
  title={On the dangers of stochastic parrots: Can language models be too big?},
  author={Bender, Emily M and Gebru, Timnit and McMillan-Major, Angelina and Shmitchell, Shmargaret},
  booktitle={Proceedings of the ACM conference on fairness, accountability, and transparency},
  pages={610--623},
  year={2021}
}

@article{cui2024risk,
  title={Risk taxonomy, mitigation, and assessment benchmarks of large language model systems},
  author={Cui, Tianyu and Wang, Yanling and Fu, Chuanpu and Xiao, Yong and Li, Sijia and Deng, Xinhao and Liu, Yunpeng and Zhang, Qinglin and Qiu, Ziyi and Li, Peiyang and others},
  journal={arXiv preprint arXiv:2401.05778},
  year={2024}
}

@article{ganguli2022red,
  title={Red teaming language models to reduce harms: Methods, scaling behaviors, and lessons learned},
  author={Ganguli, Deep and Lovitt, Liane and Kernion, Jackson and Askell, Amanda and Bai, Yuntao and Kadavath, Saurav and Mann, Ben and Perez, Ethan and Schiefer, Nicholas and Ndousse, Kamal and others},
  journal={arXiv preprint arXiv:2209.07858},
  year={2022}
}

@article{rai2024guardian,
  title={Guardian: A multi-tiered defense architecture for thwarting prompt injection attacks on llms},
  author={Rai, Parijat and Sood, Saumil and Madisetti, Vijay K and Bahga, Arshdeep},
  journal={Journal of Software Engineering and Applications},
  volume={17},
  number={1},
  pages={43--68},
  year={2024},
  publisher={Scientific Research Publishing}
}

@article{derczynski2024garak,
  title={garak: A framework for security probing large language models},
  author={Derczynski, Leon and Galinkin, Erick and Martin, Jeffrey and Majumdar, Subho and Inie, Nanna},
  journal={arXiv preprint arXiv:2406.11036},
  year={2024}
}

@article{shi2024challenges,
  title={Challenges and Exploration of Data Privacy Protection in Large Language Models},
  author={Shi, Min and Yang, Haijun},
  journal={Big Data Research},
  volume={10},
  number={5},
  pages={168--176},
  year={2024}
}

@article{meng2024large,
  title={A large-scale privacy assessment of Android third-party sdks},
  author={Meng, Mark Huasong and Yan, Chuan and Hao, Yun and Zhang, Qing and Wang, Zeyu and Wang, Kailong and Teo, Sin Gee and Bai, Guangdong and Dong, Jin Song},
  journal={arXiv preprint arXiv:2409.10411},
  year={2024}
}

@article{edge2024local,
  title={From local to global: A graph rag approach to query-focused summarization},
  author={Edge, Darren and Trinh, Ha and Cheng, Newman and Bradley, Joshua and Chao, Alex and Mody, Apurva and Truitt, Steven and Metropolitansky, Dasha and Ness, Robert Osazuwa and Larson, Jonathan},
  journal={arXiv preprint arXiv:2404.16130},
  year={2024}
}

@misc{o2023using,
  title={Using large language models to write theses and dissertations},
  author={O'Leary, Daniel E},
  journal={Intelligent Systems in Accounting, Finance and Management},
  volume={30},
  number={4},
  pages={228--234},
  year={2023},
  publisher={Wiley Online Library}
}

@article{minaee2024large,
  title={Large language models: A survey},
  author={Minaee, Shervin and Mikolov, Tomas and Nikzad, Narjes and Chenaghlu, Meysam and Socher, Richard and Amatriain, Xavier and Gao, Jianfeng},
  journal={arXiv preprint arXiv:2402.06196},
  year={2024}
}

@phdthesis{john2025owasp,
  title={Owasp top 10 for llm apps \& gen ai agentic security initiative},
  author={John, Sotiropoulos and Del, Rosario Ron F and Evgeniy, Kokuykin and Helen, Oakley and Idan, Habler and Kayla, Underkoffler and Ken, Huang and Peter, Steffensen and Rakshith, Aralimatti and Ron, Bitton and others},
  year={2025},
  school={OWASP}
}

@inproceedings{kumar2024strengthening,
  title={Strengthening LLM trust boundaries: a survey of prompt injection attacks},
  author={Kumar, Surender Suresh and Cummings, M and Stimpson, Alexander},
  booktitle={IEEE 4th International Conference on Human-Machine Systems (ICHMS)},
  pages={1--6},
  year={2024}
}

@article{mou2024research,
  title={Research Progress on Security and Privacy Protection Technologies for Large Language Models},
  author={Mou, Yiyang and Chen, Hanxiao and Li, Hongwei},
  journal={Journal of Cyberspace Security Science},
  volume={2},
  number={1},
  pages={40--49},
  year={2024},
  doi={10.20172/j.issn.2097-3136.240103}
}

@article{tai2025survey,
  title={A Survey on Adversarial Attacks and Defenses for Large Language Models},
  author={Tai, Jianwei and Yang, Shuangning and Wang, Jiajia and others},
  journal={Journal of Computer Research and Development},
  volume={62},
  number={3},
  pages={563--588},
  year={2025}
}

@article{wang2024multi,
  title={Multi-Perspective View on Large Model Security and Practice},
  author={Wang, Xiaochen and Zhang, Kun and Zhang, Peng},
  journal={Journal of Computer Research and Development},
  volume={61},
  number={5},
  pages={1104--1112},
  year={2024}
}

@article{nath2022new,
  title={New meaning for NLP: the trials and tribulations of natural language processing with GPT-3 in ophthalmology},
  author={Nath, Siddharth and Marie, Abdullah and Ellershaw, Simon and Korot, Edward and Keane, Pearse A},
  journal={British Journal of Ophthalmology},
  volume={106},
  number={7},
  pages={889--892},
  year={2022},
  publisher={BMJ Publishing Group Ltd}
}

@article{liang2024survey,
  title={A Survey on Jailbreak Attacks and Defenses for Large Language Models},
  author={Liang, Siyuan and He, Yingzhe and Liu, Aishan and others},
  journal={Journal of Information Security},
  volume={9},
  number={5},
  pages={56--86},
  year={2024}
}

@article{gallifant2024peer,
  title={Peer review of GPT-4 technical report and systems card},
  author={Gallifant, Jack and Fiske, Amelia and Levites Strekalova, Yulia A and Osorio-Valencia, Juan S and Parke, Rachael and Mwavu, Rogers and Martinez, Nicole and Gichoya, Judy Wawira and Ghassemi, Marzyeh and Demner-Fushman, Dina and others},
  journal={PLOS digital health},
  volume={3},
  number={1},
  pages={e0000417},
  year={2024},
  publisher={Public Library of Science San Francisco, CA USA}
}

@article{mao2024context,
  title={Context-aware robust fine-tuning},
  author={Mao, Xiaofeng and Chen, Yufeng and Jia, Xiaojun and Zhang, Rong and Xue, Hui and Li, Zhao},
  journal={International Journal of Computer Vision},
  volume={132},
  number={5},
  pages={1685--1700},
  year={2024},
  publisher={Springer}
}

@article{ziegler2019fine,
  title={Fine-tuning language models from human preferences},
  author={Ziegler, Daniel M and Stiennon, Nisan and Wu, Jeffrey and Brown, Tom B and Radford, Alec and Amodei, Dario and Christiano, Paul and Irving, Geoffrey},
  journal={arXiv preprint arXiv:1909.08593},
  year={2019}
}

@article{wang2024security,
  title={Security Analysis of Large Model Content Generation Based on Knowledge Editing},
  author={Wang, Mengru and Yao, Yunzhi and Xi, Zekun and others},
  journal={Journal of Computer Research and Development},
  volume={61},
  number={5},
  pages={1143--1155},
  year={2024}
}

@article{huang2022towards,
  title={Towards reasoning in large language models: A survey},
  author={Huang, Jie and Chang, Kevin Chen-Chuan},
  journal={arXiv preprint arXiv:2212.10403},
  year={2022}
}

@article{wu2024mobilevlm,
  title={Mobilevlm: A vision-language model for better intra-and inter-ui understanding},
  author={Wu, Qinzhuo and Xu, Weikai and Liu, Wei and Tan, Tao and Liu, Jianfeng and Li, Ang and Luan, Jian and Wang, Bin and Shang, Shuo},
  journal={arXiv preprint arXiv:2409.14818},
  year={2024}
}

@article{feng2024prompt,
  title={Prompt-based learning framework for zero-shot cross-lingual text classification},
  author={Feng, Kai and Huang, Lan and Wang, Kangping and Wei, Wei and Zhang, Rui},
  journal={Engineering Applications of Artificial Intelligence},
  volume={133},
  pages={108481},
  year={2024},
  publisher={Elsevier}
}

@article{liang2022adversarial,
  title={Adversarial attack and defense: A survey},
  author={Liang, Hongshuo and He, Erlu and Zhao, Yangyang and Jia, Zhe and Li, Hao},
  journal={Electronics},
  volume={11},
  number={8},
  pages={1283},
  year={2022},
  publisher={MDPI}
}

@article{wei2022emergent,
  title={Emergent abilities of large language models},
  author={Wei, Jason and Tay, Yi and Bommasani, Rishi and Raffel, Colin and Zoph, Barret and Borgeaud, Sebastian and Yogatama, Dani and Bosma, Maarten and Zhou, Denny and Metzler, Donald and others},
  journal={arXiv preprint arXiv:2206.07682},
  year={2022}
}

@article{vaswani2017attention,
  title={Attention is all you need},
  author={Vaswani, Ashish and Shazeer, Noam and Parmar, Niki and Uszkoreit, Jakob and Jones, Llion and Gomez, Aidan N and Kaiser, {\L}ukasz and Polosukhin, Illia},
  journal={Advances in neural information processing systems},
  volume={30},
  year={2017}
}

@article{brown2020language,
  title={Language models are few-shot learners},
  author={Brown, Tom and Mann, Benjamin and Ryder, Nick and Subbiah, Melanie and Kaplan, Jared D and Dhariwal, Prafulla and Neelakantan, Arvind and Shyam, Pranav and Sastry, Girish and Askell, Amanda and others},
  journal={Advances in neural information processing systems},
  volume={33},
  pages={1877--1901},
  year={2020}
}

@inproceedings{carlini2022quantifying,
  title={Quantifying memorization across neural language models},
  author={Carlini, Nicholas and Ippolito, Daphne and Jagielski, Matthew and Lee, Katherine and Tramer, Florian and Zhang, Chiyuan},
  booktitle={The Eleventh International Conference on Learning Representations},
  year={2022}
}

@inproceedings{fredrikson2015model,
  title={Model inversion attacks that exploit confidence information and basic countermeasures},
  author={Fredrikson, Matt and Jha, Somesh and Ristenpart, Thomas},
  booktitle={Proceedings of the 22nd ACM SIGSAC conference on computer and communications security},
  pages={1322--1333},
  year={2015}
}

@article{bolukbasi2016man,
  title={Man is to computer programmer as woman is to homemaker? debiasing word embeddings},
  author={Bolukbasi, Tolga and Chang, Kai-Wei and Zou, James Y and Saligrama, Venkatesh and Kalai, Adam T},
  journal={Advances in neural information processing systems},
  volume={29},
  year={2016}
}

@inproceedings{pu2023deepfake,
  title={Deepfake text detection: Limitations and opportunities},
  author={Pu, Jiameng and Sarwar, Zain and Abdullah, Sifat Muhammad and Rehman, Abdullah and Kim, Yoonjin and Bhattacharya, Parantapa and Javed, Mobin and Viswanath, Bimal},
  booktitle={2023 IEEE symposium on security and privacy (SP)},
  pages={1613--1630},
  year={2023},
  organization={IEEE}
}

@inproceedings{schmidt2017survey,
  title={A survey on hate speech detection using natural language processing},
  author={Schmidt, Anna and Wiegand, Michael},
  booktitle={Proceedings of the fifth international workshop on natural language processing for social media},
  pages={1--10},
  year={2017}
}

@inproceedings{pennington2014glove,
  title={Glove: Global vectors for word representation},
  author={Pennington, Jeffrey and Socher, Richard and Manning, Christopher D},
  booktitle={Proceedings of the 2014 conference on empirical methods in natural language processing (EMNLP)},
  pages={1532--1543},
  year={2014}
}

@article{shannon1948mathematical,
  title={A mathematical theory of communication},
  author={Shannon, Claude E},
  journal={The Bell system technical journal},
  volume={27},
  number={3},
  pages={379--423},
  year={1948},
  publisher={Nokia Bell Labs}
}

@inproceedings{devlin2019bert,
  title={Bert: Pre-training of deep bidirectional transformers for language understanding},
  author={Devlin, Jacob and Chang, Ming-Wei and Lee, Kenton and Toutanova, Kristina},
  booktitle={Proceedings of the 2019 conference of the North American chapter of the association for computational linguistics: human language technologies, volume 1 (long and short papers)},
  pages={4171--4186},
  year={2019}
}

@inproceedings{rajbhandari2020zero,
  title={Zero: Memory optimizations toward training trillion parameter models},
  author={Rajbhandari, Samyam and Rasley, Jeff and Ruwase, Olatunji and He, Yuxiong},
  booktitle={SC20: International Conference for High Performance Computing, Networking, Storage and Analysis},
  pages={1--16},
  year={2020},
  organization={IEEE}
}

@inproceedings{abadi2016deep,
  title={Deep learning with differential privacy},
  author={Abadi, Martin and Chu, Andy and Goodfellow, Ian and McMahan, H Brendan and Mironov, Ilya and Talwar, Kunal and Zhang, Li},
  booktitle={Proceedings of the 2016 ACM SIGSAC conference on computer and communications security},
  pages={308--318},
  year={2016}
}

@inproceedings{kirchenbauer2023watermark,
  title={A watermark for large language models},
  author={Kirchenbauer, John and Geiping, Jonas and Wen, Yuxin and Katz, Jonathan and Miers, Ian and Goldstein, Tom},
  booktitle={International Conference on Machine Learning},
  pages={17061--17084},
  year={2023},
  organization={PMLR}
}

@inproceedings{papineni2002bleu,
  title={Bleu: a method for automatic evaluation of machine translation},
  author={Papineni, Kishore and Roukos, Salim and Ward, Todd and Zhu, Wei-Jing},
  booktitle={Proceedings of the 40th annual meeting of the Association for Computational Linguistics},
  pages={311--318},
  year={2002}
}

@inproceedings{lecun20191,
  title={1.1 deep learning hardware: Past, present, and future},
  author={LeCun, Yann},
  booktitle={2019 IEEE International Solid-State Circuits Conference-(ISSCC)},
  pages={12--19},
  year={2019},
  organization={IEEE}
}

\end{document}